\documentclass[preprint]{aastex631}

\begin{document}

\title{The First Y Dwarf Data From JWST Show That Dynamic and Diabatic Processes Regulate Cold Brown Dwarf Atmospheres}

\author[0000-0002-3681-2989]{S. K. Leggett}
\affiliation{Gemini Observatory/NSF's NOIRLab, 670 N. A'ohoku Place, Hilo, HI 96720, USA}

\author[0000-0001-6172-3403]{Pascal Tremblin}
\affiliation{Universite Paris-Saclay, UVSQ, CNRS, CEA, Maison de la Simulation, 91191, Gif-sur-Yvette, France
}



\begin{abstract}

The {\em James Webb Space Telescope (JWST)} is now observing Y dwarfs, the coldest known brown dwarfs, with effective temperatures $T_{\rm eff} \lesssim 475$~K. The first published observations provide important information: not only is the atmospheric chemistry out of equilibrium, as previously known, but the pressure-temperature profile is not in the standard adiabatic form. The rapid rotation of these Jupiter-size, isolated, brown dwarfs dominates the atmospheric dynamics, and thermal and compositional changes disrupt convection. These processes produce a  colder lower atmosphere, and a warmer upper atmosphere,  compared to a standard adiabatic profile. \citet{Leggett_2021} presented  empirical models where the pressure-temperature profile was adjusted so that synthetic spectra reproduced the $1 \lesssim \lambda~\mu$m $\lesssim 20$ spectral energy distributions of brown dwarfs with  $260 \leq T_{\rm eff}$~K $\leq 540$.
We show that spectra generated by these models fit the first {\em JWST} Y dwarf spectrum better than standard-adiabat models.  
Unexpectedly,  there is no $4.3~\mu$m PH$_3$ feature in the {\em JWST} spectrum and
atmospheres without phosphorus better reproduce the $4~\mu$m flux peak.
Our analysis of new {\em JWST} photometry indicates that 
the recently discovered faint secondary of the  WISE J033605.05-014350AB system
\citep{Calissendorff_2023} has $T_{\rm eff} \approx 295$~K, making it the first dwarf in the significant luminosity gap between the 260~K WISE J085510.83-071442.5, and all other known Y dwarfs.
The adiabat-adjusted disequilibrium-chemistry models are
recommended for analyses of 
all brown dwarfs cooler than 600~K, and a grid is publicly available. Photometric color transformations
are provided in an Appendix.

\end{abstract}




\medskip
\section{Introduction} \label{sec:intro}

One hundred years ago, the 5~pc region of space with our Sun at its center was known to include three M dwarfs with luminosities around $10^{-3} L_{\odot}$ --- Barnard's star,  Proxima Centauri, and Wolf 359 \citep{Barnard_1916, Voute_1917, Wolf_1919}. Improvements in infrared detector technology in the 1980s enabled the detection of colder sources which are fainter by 5 orders of magnitude. Currently, we know that same 5~pc region also contains an L, a T and a Y dwarf \citep{Luhman_2013, Luhman_2014}. The effective temperatures ($T_{\rm eff}$) of the known L, T, and Y dwarfs are approximately 2000 -- 1200~K, 1200 -- 475~K, and 475 -- 250~K respectively \citep[e.g.][]{Stephens_2009, Kirkpatrick_2021, Leggett_2021}. Many advances have been made in modelling the atmospheres of these objects, and hence their emitted spectral energy distributions (SEDs). Processes such as condensation of the refractory elements and grain sedimentation, and disequilibrium chemistry driven by vertical transport of gas, are now known to be critical, changing the abundance of dominant species by orders of magnitude \citep[e.g.][]{Fegley_1996, Griffith_1999, Lodders_1999, Ackerman_2001, Allard_2001, Saumon_2006, Leggett_2016a}.  The SEDs of cool stars and brown dwarfs -- objects with masses too low for sustained fusion
\citep[e.g.][]{Burrows_1993, Saumon_2008, Phillips_2020}
-- can now be well reproduced, at least for  $T_{\rm eff} > 600$~K \citep[e.g.][]{Stephens_2009, Leggett_2017}.

During the last three years new models of cold dwarf atmospheres have become available. \citet{Karalidi_2021} and \citet{Marley_2021} present Sonora brown dwarf atmospheric and evolutionary models for cloud-free atmospheres with a range of metallicity, both in and out of chemical equilibrium. 
\citet{Lacy_2023} present models of Y dwarfs with water clouds and disequilibrium chemistry, with a range of metallicity.  The latter models calculate spectra which closely agree with the Sonora spectra, although the pressure-temperature relationships in the upper atmosphere differ 
\citep[Figures 20 and 22 of][]{Lacy_2023}. We note that the primary photospheric region of a Y dwarf is expected to be cloud free, as condensation curves of alkalis, sulfides, and chlorides show that those grains would exist deep in the atmosphere, and similarly water clouds would only exist high in the atmosphere \citep[e.g.][]{Morley_2014, Leggett_2021}.

Researchers have also approached the analysis of atmospheres using the retrieval method, where observations guide the generation of the model structure. \citet{Zalesky_2019, Zalesky_2022} and \citet{Hood_2023} adopt this approach in analyses of T and Y dwarfs. To date these analyses have been limited to the near-infrared spectral region only, meaning that a limited region of the photosphere is probed. Generally the retrieval analyses determine atmospheric parameters that are consistent with the forward model grid analyses, however in some cases non-physical values emerge from the retrieval analysis, suggesting an incomplete understanding of the atmosphere \citep[e.g.][]{Hood_2023}.

For Y dwarfs, most model spectra significantly underestimate the flux at wavelengths between $2~\mu$m and $4~\mu$m. For example, the synthetic Sonora-Cholla colors presented by \citet{Karalidi_2021} have $K$ magnitudes which are about a magnitude too faint, unless the surface gravity is reduced to an unlikely, very low, value 
\citep[Figure 18 of][]{Karalidi_2021}. Similarly the synthetic spectra generated by the models of \citet{Lacy_2023} are a factor of two to three too faint between $2~\mu$m and $3.8~\mu$m
\citep[Figure 17 of][]{Lacy_2023}.
\citet{Leggett_2021} demonstrate that in order to reproduce both the near-infrared and mid-infrared spectra of Y dwarfs, the lower atmosphere must be cooler and the upper atmosphere warmer, compared to a standard pressure-temperature adiabatic relationship. 
Deviations from the standard adiabat are 
a natural outcome of rapid rotation \citep{Tan_2021}
and thermal and compositional changes in the atmosphere \citep{Tremblin_2019}.

In Section 2 we briefly describe the ATMO 2020 models 
\citep{Phillips_2020} which were empirically adjusted for \citet{Leggett_2021} and extended into a larger grid for \citet{Meisner_2023}.  
We refer to the latter model grid set as the ATMO2020++ models.
In Section 3 we compare 
the first published {\em JWST} spectrum of a Y dwarf, and its best-fit Sonora model SED, to an ATMO2020++ SED.
In Section 4 we present an analysis of color-color diagrams using recently published {\em JWST} photometry for four Y dwarfs. Section 5 presents our conclusions. 

The Appendices supply supporting material. Appendix A adds more detailed information about the ATMO2020++ atmospheres. Appendix B presents a comparison of the Y dwarf spectrum to spectra generated by other model grids (BT-Settl \citep{Allard_2014, Allard_2016} and \citet{Lacy_2023}), for reference. Appendix C provides
transformations between ground-based near-infrared, {\em Spitzer} \citep{Werner_2004}, and {\em Wide-field Infrared Survey Explorer} 
\citep[{\em WISE},][]{Wright_2010} colors and {\em JWST} colors.

\medskip
\section{Pressure-Temperature Profiles 
and the ATMO2020++ Models} 

One-dimensional models, such as the ATMO and Sonora models, represent the atmosphere as a pressure ($P$) - temperature ($T$) profile that maps the cooling from the core out to the surface, and by a chemical abundance profile that maps the chemical changes that occur through the atmosphere as $P$ and $T$ change. The $P$ - $T$ profile can be thought of as a slice through the atmosphere, where both temperature and pressure decrease with increasing altitude. Energy transport in a cool dwarf atmosphere is predominantly convective, with radiative cooling becoming important high in the atmosphere 
where the pressure is too low for convection to be efficient. Convection is treated as an adiabatic process where ${P}^{(1-\gamma )}{T}^{\gamma }=\mathrm{constant}$. 
For an ideal gas, $\gamma$ is the ratio of specific heats at constant pressure and volume and, for a gas composed entirely of molecular hydrogen, $\gamma = 1.4$ \citep[see][for reviews of atmospheric processes]{Marley_2015, Zhang_2020}.

However isolated brown dwarfs rotate rapidly, with periods of a few hours, as demonstrated by rotational line broadening 
\citep[e.g.][]{Hsu_2021} and inferred from observed variability
\citep[e.g.][]{Metchev_2015, Miles_2017, Tannock_2021}, and rotation strongly modulates the heat transport from the interior of the brown dwarf. Three-dimensional simulations of turbulent, convective, rotating atmospheres produce vertical velocities consistent with observed chemical mixing, and surface features consistent with observed variability \citep{Showman_2013}. Simulations also show that rotation dramatically changes the $P$ - $T$ profile. For example, 
\citet{Tan_2021} explore the  profile of a brown dwarf atmosphere with and without rotation, and find that the rotating cloud-free atmosphere is significantly cooler in the lower atmosphere,  and has an almost isothermal upper atmosphere (their Figure 4). The three-dimensional models also show that the surface of rotating brown dwarfs varies with latitude \citep{Showman_2019}, suggesting that the observed properties of brown dwarfs are dependent on both rotational speed and axis inclination \citep{Lipatov_2022}.

The turbulent convective atmospheres of brown dwarfs are also likely to be subject to diabatic processes, as described by \citet{Tremblin_2019} \citep[see also][]{Tan_2017}. These could include compositional changes of carbon and nitrogen bearing molecules (CO/CH$_4$, N$_2$/NH$_3$), and condensation of the alkalis, chlorides and sulfides \citep[e.g.][]{Morley_2012, Morley_2014}.

In \citet{Leggett_2021} we used the ATMO2020  disequilibrium models \citep{Phillips_2020} as a starting point, and
treated the adiabatic parameter $\gamma$ as a variable, along with $T_{\rm eff}$, surface gravity $g$, metallicity [m/H], and the diffusion coefficient $K_{zz}$ which characterizes the transport of gas and resulting disequilibrium chemistry. 
The adiabat parameter $\gamma$  was set to the standard value (i.e. $\approx 1.4$) at a depth in the atmosphere defined by  pressure $P(\gamma ,\max )$ bar, and deeper, and reduced to a constant, smaller, value higher in the atmosphere. This approach was tested against the SEDs of one late-T dwarf and six Y dwarfs. By tuning $T_{\rm eff}$, $g$, [m/H], $K_{zz}$, $\gamma$, and $P(\gamma ,\max )$ for each brown dwarf, 
good fits across the entire SED were obtained. A limited grid of models with $K_{zz} = 10^7$ cm$^2$ s$^{-1}$, $\gamma = 1.25$, and $P(\gamma ,\max ) = 15$~bar was generated, which demonstrated significant and comprehensive improvements in the agreement between modeled and observed colors for a large sample of late-T and Y dwarfs.
 
Following \citet{Leggett_2021}, a larger grid of adiabat-adjusted models was generated for a study of metal-poor brown dwarfs by \citet{Meisner_2023}.  We refer to this model set as the ATMO2020++ models. Appendix A gives more information about these atmospheres, including how the parameters were generalized from the smaller to the larger grid, and how the resulting synthetic SEDs compare.  The ATMO2020++ models generate colors which agree well with observations, as shown in Figures 2 and 3 of \citet{Meisner_2023}.  We show below that the ATMO2020++ models perform extremely well when compared to the continuous $1 \leq \lambda~\mu$m $\leq 12$ SED measured by {\em JWST} for a Y0 dwarf. 
ATMO2020++ spectra are available at:\dataset[the ERC-ATMO Opendata page]
{https://noctis.erc-atmo.eu/fsdownload/Q7MUSoCLR/meisner2023}.

\bigskip
\section{The First JWST Y Dwarf Spectrum} 

\begin{figure}[ht!]
\plotone{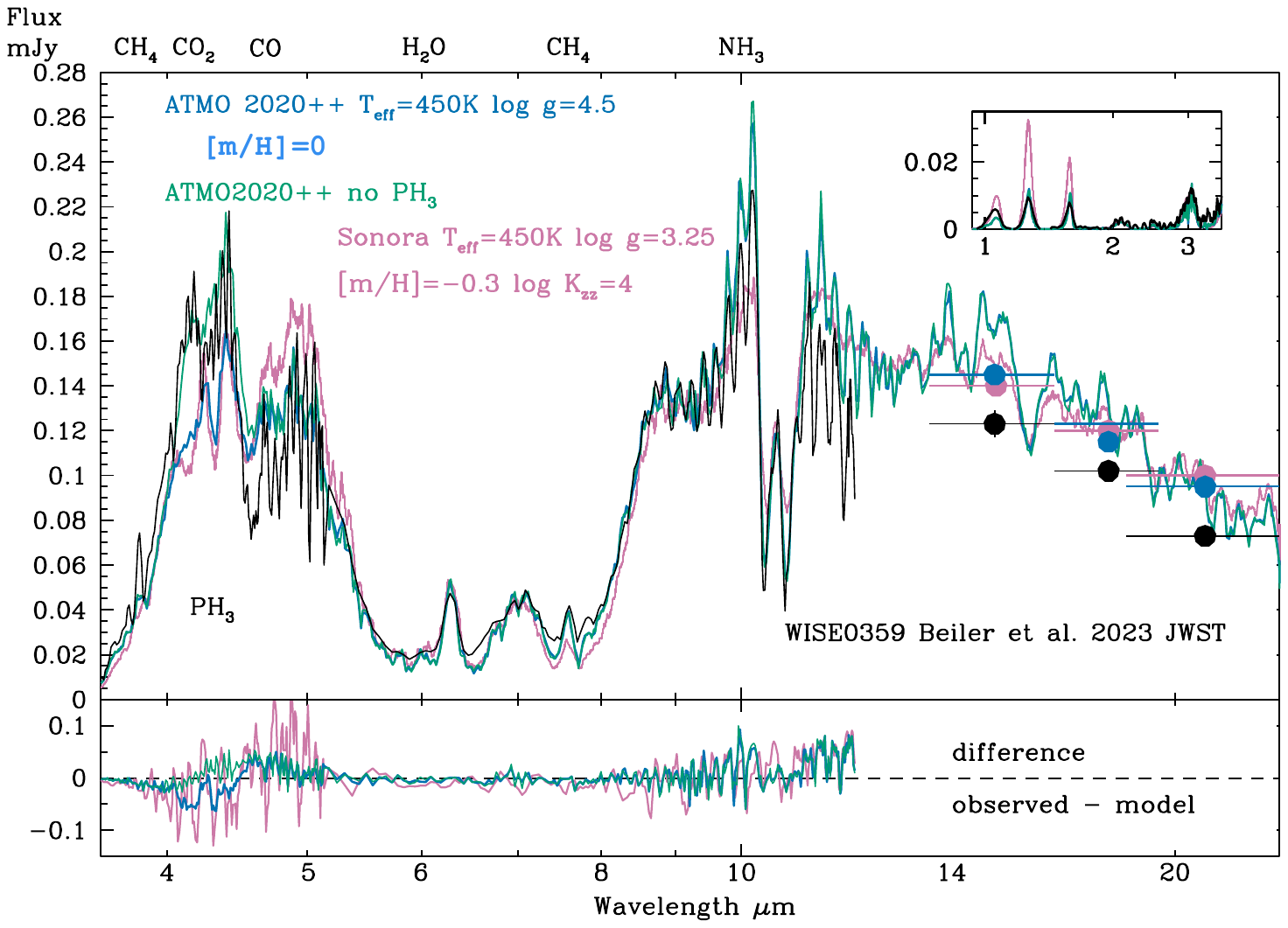}
\vskip -0.2in
\caption{The black line is the observed spectrum of  WISE 035934.06-540154.6 from \citet{Beiler_2023}. Black points represent their photometric measurements. The pink line and dots are the best fit Sonora model spectrum and photometry, from Beiler et al. The blue line and data points are the spectrum and photometry from the best fit model of the ATMO2020++ grid.  The green line is the same model where PH$_3$ has been removed from the atmospheric composition. The ATMO2020++ model is scaled by the known distance to WISE 0359 \citep{Kirkpatrick_2021} and the radius inferred from evolutionary models \citep{Marley_2021}, and has not been scaled to fit.
Principal opacity species are indicated. The lower panel shows the difference between the observed spectrum and the spectra generated by the  three models (note the difference in $y$-axis scale in the upper and lower panels). The inset shows the near-infrared flux region. See Section 3 for discussion.
}
\end{figure}

The cryogenic portion of the {\em Spitzer} mission ended in 2009. The  {\em Spitzer} infrared spectrograph \citep{Houck_2004} provided exquisite mid-infrared spectra of T dwarfs as cool as $T_{\rm eff} \approx 600$~K \citep[e.g.][]{Cushing_2006, Leggett_2009}. Since that date, only limited spectra have been obtained at wavelengths longer than $\lambda \approx 2.5~\mu$m, wavelengths where cool brown dwarfs emit significant flux. While progress has been made using the mid-infrared photometric data from {\em Spitzer} and {\em WISE}, and some $\lambda \approx 4~\mu$m data from the ground, the community has been anxiously awaiting the first {\em JWST} data for Y dwarfs \citep{Marley_2009}.

\citet{Beiler_2023} published the first {\em JWST} spectrum of a Y dwarf, together with longer wavelength photometry. The object is WISE J035934.06-540154.6 (herafter WISE 0359), which is classified as a Y0 dwarf \citep{Kirkpatrick_2021} and appears to be typical of its type (see Section 4). Figure 1 shows the observational data, with the best fit Sonora spectrum from Beiler et al., and a fit using the ATMO 2020++ models. 
A fit with a disequilibrium metal-poor Sonora model reproduces the observations quite well (Figure 1) however the required low surface gravity implies an unlikely very young age and low mass \citep{Beiler_2023}.

We find that an ``off-the-shelf'' ATMO 2020++ model with $T_{\rm eff} = 450$~K, log $g = 4.5$, and [m/H] $= 0$
produces an excellent fit to the data, as shown in Figure 1.
An increase or decrease in $T_{\rm eff}$ by 50~K changes the luminosity significantly, and the flux levels at $4.5~\mu$m and $12.0~\mu$m would increase or decrease by 40\%, meaning the 400~K and 500~K models do not match the observations. Figure 6 of \citet{Leggett_2021} illustrates the dependency of the SED on the other atmospheric parameters. For this observed spectrum, we found that the relative heights of the $4.5~\mu$m and $10~\mu$m  flux peaks, the height and shape of the 8 -- 9~$\mu$m shoulder, and the depths of the CO and NH$_3$ absorption features, allowed us to constrain log $g$ and [m/H], and to select the best fit from our available grid by eye. 

The near-infrared spectral region was not considered in the fitting process, because very little flux emerges there ($5\%$ of the total flux).  However Figure 1 shows that the ATMO2020++ model produces an excellent match to the near-infrared observations. For reference, Appendix B presents a comparison of the {\em JWST} spectrum to a 450~K BT-Settl model spectrum \citep{Allard_2014, Allard_2016} as the BT-Settl grid is commonly used for warmer brown dwarfs. The BT-Settl model produces an inferior fit compared to Sonora or ATMO2020++, at these temperatures. Appendix B also compares the spectrum to a 450~K disequilibrium cloud-free spectrum from \citet{Lacy_2023}. The agreement is again inferior, although we find in the next Section that the \citet{Lacy_2023} 
disequilibrium spectra which include water clouds may agree better with the observations of Y dwarfs cooler than 350~K (Figure 2).

\citet{Beiler_2023} determine $F_{\rm bol} = 6.89 \pm 0.16 \times 10^{-17}$ W m$^{-2}$ for WISE 0359, by integrating over the observed 1 -- 12~$\mu$m spectrum, applying a linear interpolation over the 12 -- 21~$\mu$m photometric data points, and adopting a Rayleigh-Jean tail for the longest wavelengths. Our adopted ATMO2020++ model integrates to 
$7.41 \times 10^{-17}$ W m$^{-2}$, and is within  $\approx 3~\sigma$ of the measured value.
The over-luminosity seen in Figure 1 for our model at  $14 \lesssim \lambda~\mu$m $\lesssim 21$ likely contributes to the difference, however we note that, at this early stage in the {\em JWST} pipeline, small errors in the MIRI flux calibration cannot be excluded and these may contribute to the apparent over-luminosity.
The models indicates that a Rayleigh-Jean tail is a good approximation for $\lambda > 21~\mu$m.  

\citet{Beiler_2023} use the bolometric luminosity and evolutionary models to determine $T_{\rm eff} = 467^{+16}_{-18}$~K. The $T_{\rm eff}$ 
for both the Sonora and ATMO2020++ model fits agree with this value, within the uncertainties. However the model parameters differ significantly in other ways, and the ATMO2020++ model provides a superior fit as can be seen in Figure 1. Key differences are:
\begin{itemize}
\item \underline {Metallicity:} 
The colors of WISE 0359 indicate that it is a typical field object (Section 4). A solar metallicity would therefore be expected, as we find here, compared to the [m/H] $= -0.3$~dex determined by \citet{Beiler_2023}.
\item \underline {Surface Gravity:}
The ATMO2020++  model parameters of $T_{\rm eff} = 450$~K and log $g = 4.5$, together with  evolutionary models \citep{Marley_2021}, give an age and mass for WISE 0359 of around 2.5~Gyr and 14 Jupiter masses. This compares to 20~Myr and 1 Jupiter mass for the  lower gravity \citet{Beiler_2023} Sonora model. The kinematic age of the local T dwarf population is around 3.5~Gyr \citep{Hsu_2021}, and simulations of the field substellar population show a median age for a 450~K Y dwarf of around 6~Gyr \citep{Kirkpatrick_2021}. Hence a value of log $g = 4.5$, with an associated age of a few Gyr, is more plausible than a value of log $g = 3.5$, with an associated age of a few Myr.
\item \underline {Opacities:} 
The ATMO2020++ model (blue line) reproduces the spectral features seen in Figure 1 better than the Sonora model (pink line, see the difference spectra in lower panel).  In particular, the near-infrared peaks, the  $\lambda \approx 5.0~\mu$m region  where opacities due to CO and H$_2$O dominate \citep[Figure 10]{Leggett_2021}, and the strong double NH$_3$ absorption at $\lambda \approx 10.5~\mu$m are better matched.
\end{itemize}

Both the Sonora and ATMO models generate spectra with strong PH$_3$ absorption features at $4.1 \lesssim \lambda~\mu$m $\lesssim 4.3$ \citep[see e.g. Figure 10 of][]{Leggett_2021}. 
This is a disequilibrium species, with PH$_3$ expected to be the stable form of phosphorus in cool atmospheres \citep{Visscher_2006}.  However the chemical pathways for phosphorus are more complex \citep{Visscher_2006, Wang_2016}
and less certain \citep{Bains_2023}. Although the feature is seen in the solar system giant planets, it may be that the different composition or gravity of the brown dwarf atmosphere results in phosphorus taking a different form. For this work we generated a small number of ATMO2020++ model atmospheres with no phosphorus. Figure 1 shows that such a model better reproduces the $4~\mu$m flux peak for this Y0 brown dwarf. Removing this opacity does not significantly change the synthetic {\em JWST} F480M magnitude, but will brighten the {\em Spitzer} and {\em WISE} [4.5] and W2 magnitudes  by $\sim 20\%$.

\bigskip
\section{The First JWST Y Dwarf Photometry} 

\begin{figure}[ht!]
\plotone{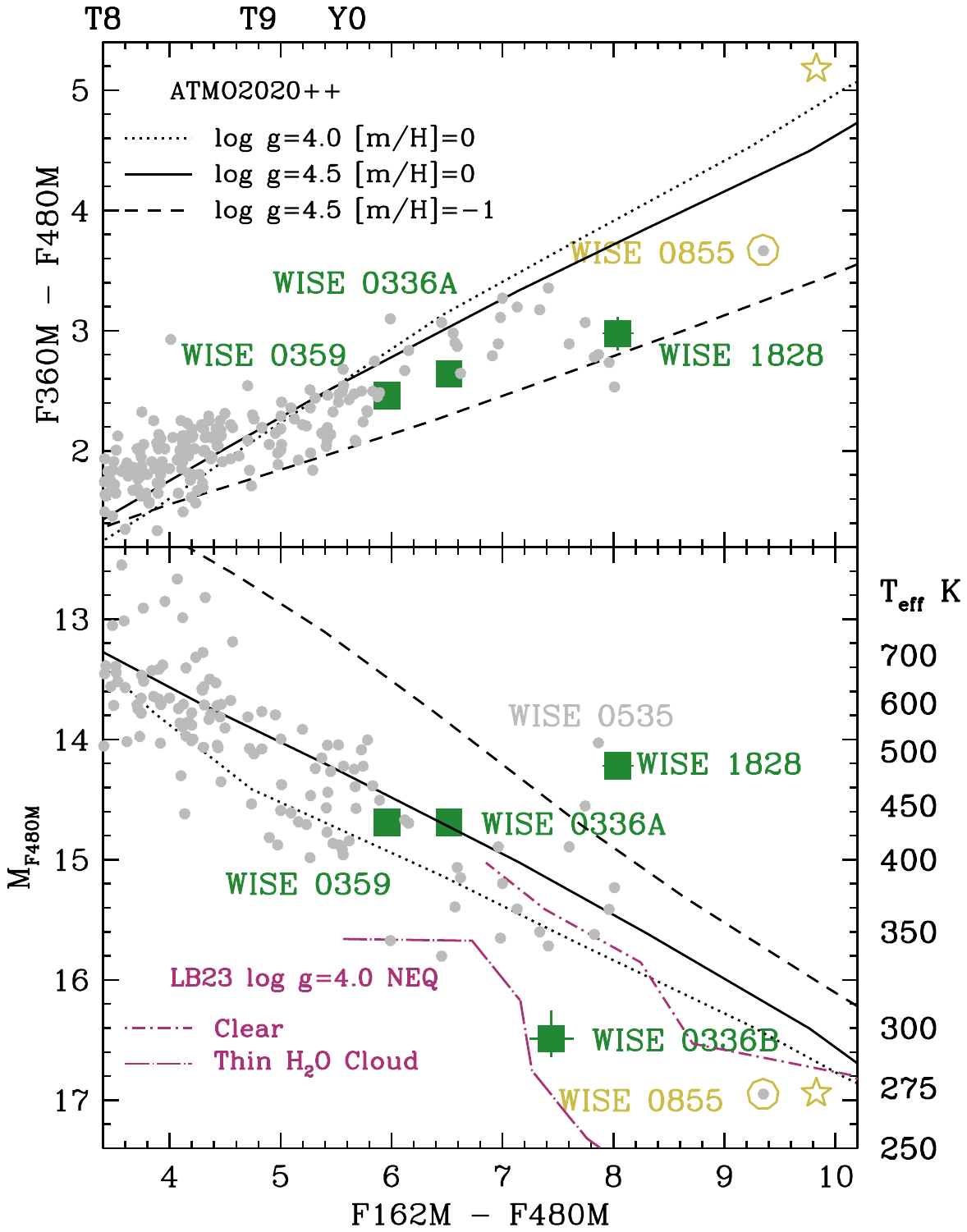}
\vskip -0.4in
\caption{Color magnitude diagrams using {\em JWST} filters expected to be commonly used for brown dwarfs. Grey dots represent the photometric sample from \citet{Leggett_2021} converted on to the {\em JWST} system.  Green squares represent recently published  {\em JWST} data. Yellow circles and stars indicate the observed and synthesized  colors, respectively, for the 260~K Y dwarf WISE 0855. 
Black lines are ATMO2020++ model sequences with parameters as given in the upper legend. Purple lines are disequilibrium chemistry sequences from \citet{Lacy_2023}, for 
$250 \leq T_{\rm eff}$~K $\leq 350$, and other
parameters as given in the lower legend. See text for further discussion. WISE 0535  (WISE J053516.80-750024.9), like WISE 1828, is an unresolved apparently overluminous system \citep[e.g.][]{Leggett_2021}.
}
\end{figure}

{\em JWST} photometry has been published for three Y dwarf systems, at the time of writing. We discuss each system separately below. To facilitate comparison of these new data with previous ground-based near-infrared, {\em Spitzer}, and {\em WISE} observations, as well as future {\em JWST} observations, we have calculated color transformations for a selection of filters using our ATMO2020++ grid. The photometry is available at: \dataset[10.5281/zenodo.7931460] {https://zenodo.org/record/7931460}. Appendix C presents color-color plots for the transformations, as well as polynomial fits for interpolation.

We have used the transformations in Appendix C to convert existing data \citep[from][]{Leggett_2021} to the {\em JWST} F162M, F360M, and F480M systems. These filters were selected because the differences $H -$ F162M, {\em Spitzer} [3.6] $-$ F360M, and {\em Spitzer} [4.5] (or {\em WISE} W2) $-$ F480M are small (see Appendix C), and therefore the transformation is less prone to error. Also, these filters are  expected to be commonly used for brown dwarf studies, and are included in current {\em JWST} programs to study Y dwarfs (e.g. GTO 1189). We note that there is no {\em JWST} filter with a bandpass similar to the ground-based $1.25~\mu$m $J$-band. The 1.0 --- 1.3~$\mu$m  F115W filter bandpass
includes strong absorption bands of H$_2$O, CH$_4$, and NH$_3$ for Y dwarfs \citep[e.g.][]{Lacy_2023}, and the filter transformations indicate that the $J -$ F115W color is very sensitive to metallicity. Historically the $J$ filter has been the default for ground-based Y dwarf followup, however the $H$ filter may now be a better choice. 

To estimate F162M from the photometry sample, the $H$ magnitude was used, unless $H$ was not available in which case $J$ was used, together with the color $H$ (or $J$) $-$ [4.5] (or W2). To estimate F360M, the [3.6] magnitude was used, unless that was not available, in which case {\em WISE} W1 was used, together with the color [3.6] $-$ [4.5] (or W1 $-$ W2). Similarly, to estimate F480M, the {\em Spitzer} [4.5] magnitude was used, unless that was not available, in which case {\em WISE} W2 was used, together with the color [3.6] $-$ [4.5] (or W1 $-$ W2).  

Figure 2 shows color-color diagrams where the green symbols are the new {\em JWST} data and smaller grey dots are the ground-based near-infrared, {\em Spitzer}, and {\em WISE} observations transformed on to the  {\em JWST} system. 
Known extreme metal-poor T dwarfs \citep[e.g.][]{Meisner_2023} have been excluded from the Figure because those color transformations are more uncertain due to the  dependency  on metallicity (see Appendix C).  In Figure 2 we show sequences for three ATMO2020++ models: solar metallicity log $g = 4.0$ and log $g = 4.5$ models, and a metal-poor 
log $g = 4.5$ model. The log $g = 4.0$ model approximately 
corresponds to an age of $0.8$~Gyr for the Y dwarfs, and the log $g = 4.5$ to $\sim 6$~Gyr  \citep{Marley_2021}.
Figure 3 and Table 1 give relationships between the F162M $-$ F480M and F360M $-$ F480M colors and $T_{\rm eff}$, and the absolute F480M magnitude  and $T_{\rm eff}$, for these three models. Table 2 gives the F162M, F360M, and F480M magnitudes for the three systems discussed below, as well as the  $T_{\rm eff}$ values inferred from the color relationships. 

F162M $-$ F480M:$M_{F480M}$ sequences from the disequilibrium chemistry models by \cite{Lacy_2023} are also shown in Figure 2 for $250 \leq T_{\rm eff}$~K $\leq 350$. The onset of water clouds reduces the emergent flux at F480M, and increases the flux at F162M \citep{Lacy_2023}. We discuss the implications of this for the two coolest objects, WISE 0336B and WISE 0855 below.


\begin{figure}[ht!]
\plotone{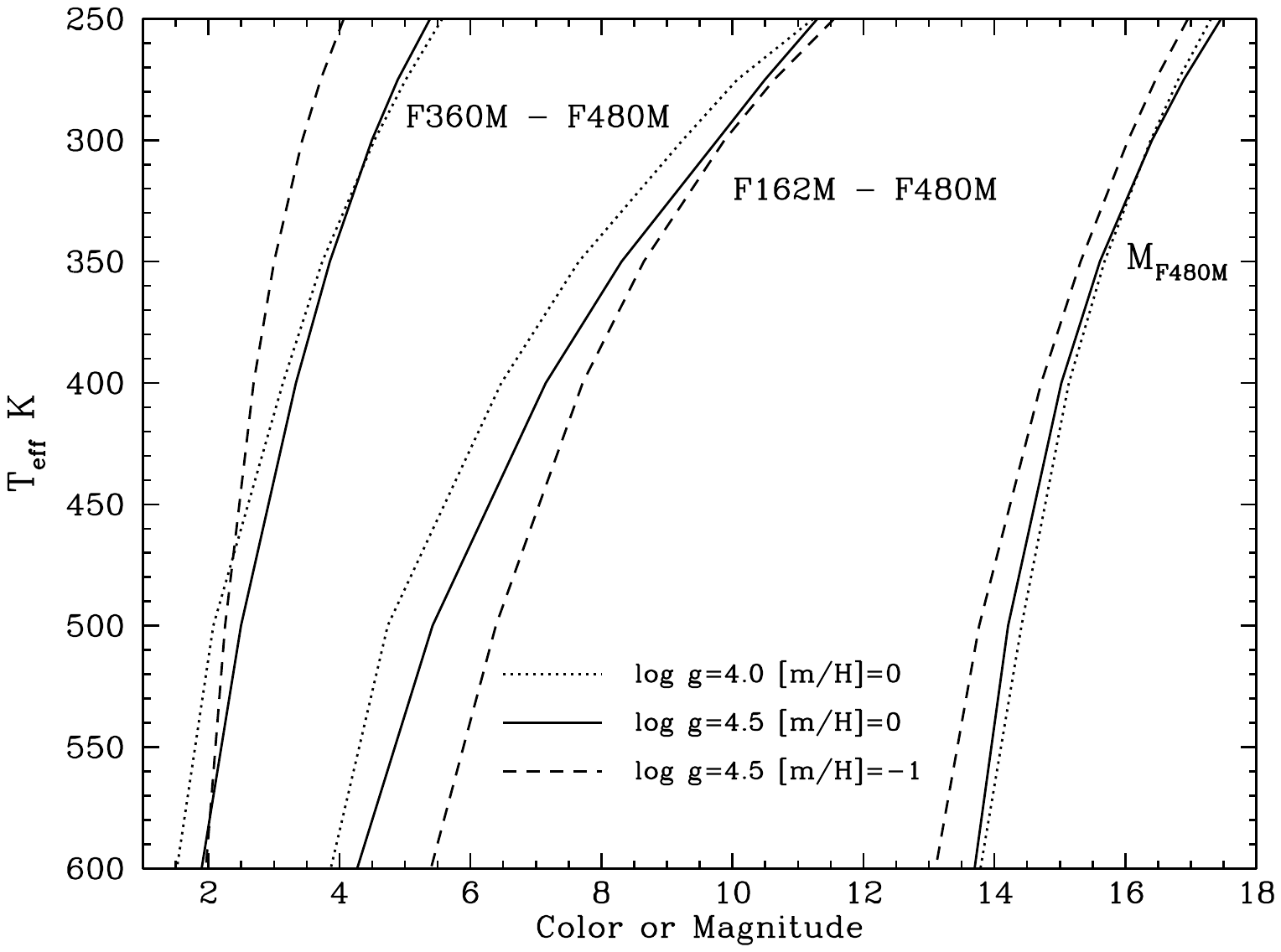}
\vskip -0.2in
\caption{Relationships between colors and $T_{\rm eff}$, calculated from the grid of ATMO2020++ colors. See also Table 1. For   $T_{\rm eff} \lesssim 350$~K the  F162M $-$ F480M and F360M $-$ F480M colors are less reliable indicators of  $T_{\rm eff}$ than $M_{F480M}$ (see Figure 2 and Section 4).
}
\end{figure}

\begin{deluxetable*}{crrrrrr}[ht!]
\tabletypesize{\normalsize}
\tablecaption{Polynomial Relationships for Estimating $T_{\rm eff}$ from Color}
\tablehead{
\colhead{Color} & \colhead{log $g$} & \colhead{$m/H$} & \colhead{$a_0$}
 & \colhead{$a_1$} & \colhead{$a_2$} & \colhead{$a_3$} 
}
\startdata
$M_{F480M}$ & 4.0 & 0.0 & +11884  &  -1694.7366 &  +81.1315 &  -1.2733  \\ 
$M_{F480M}$ & 4.5 & 0.0 & +26131 &  -4412.5066  & +252.7961 &  -4.8658 \\
$M_{F480M}$ & 4.5 & -1.0 & +41588 &  -7953.9265 &  +514.4639 &  -11.1589 \\
F162M $-$ F480M & 4.0 & 0.0 & +1700  &  -450.9727 &  +51.2187 &   -2.0184 \\
F162M $-$ F480M & 4.5 & 0.0 & +1641  & -398.6753  &  +43.3689  & -1.6876 \\
F162M $-$ F480M & 4.5 & -1.0 & +1955 &  -416.6833 &  +37.0493 &  -1.1932 \\
F360M $-$ F480M & 4.0 & 0.0 & +1134  & -474.2423 &  +98.9609  & -7.6338 \\
F360M $-$ F480M & 4.5 & 0.0 & +1249  & -488.7698  &  +91.4434  & -6.5519 \\
F360M $-$ F480M & 4.5 & -1.0 & +2717 &  -1767.2828  &  +433.6942  & -36.4072 \\
\enddata
\vskip 0.1in
\tablecomments{
$T_{\rm eff}$ is estimated using\\
$T_{\rm eff} = a_0 + a_1\times{\rm Color} + a_2\times{\rm Color}^2  + a_3\times{\rm Color}^3$\\
Excluding any systematic errors in the models, the uncertainty in   $T_{\rm eff}$ is $\pm 10$~K.
}
\end{deluxetable*}

\begin{deluxetable*}{lrllll}[ht!]
\tabletypesize{\normalsize}
\tablecaption{{\em JWST} Magnitudes and Estimated Photometric $T_{\rm eff}$ for Three Y Dwarf Systems}
\tablehead{
\colhead{{\em WISE} Name} & \colhead{Parallax$^a$} & \colhead{F162M} & \colhead{F360M} & \colhead{F480M} &  \colhead{$T_{\rm eff}^b$} \\
\colhead{RA/Dec.} & \colhead{mas} & \colhead{mag} & \colhead{mag} & \colhead{mag} &  \colhead{K} 
}
\startdata
J033605.05-014350.4A$^c$ & 99.8$\pm$ 2.1 & 21.23$\pm$ 0.10  & 17.35$\pm$ 0.07 & 14.71$_{-0.05}^{+0.02}$ & 440$\pm$ 24$^d$\\
J033605.05-014350.4B$^c$ & 99.8$\pm$ 2.1 & 23.95$_{-0.30}^{+0.18}$   &   &  16.51$_{-0.26}^{+0.12}$ & 295$\pm$ 10$^d$ \\
J035934.06-540154.6$^e$ & 73.6$\pm$ 2.0 & 21.32$\pm$ 0.05 & 17.82$\pm$ 0.05& 15.36$\pm$ 0.05 & 463$\pm$ 28$^f$\\
J182831.08+265037.8$^g$ & 100.3$\pm$ 2.0 & 22.25$\pm$ 0.10 & 17.19$\pm$ 0.10 & 14.21$\pm$ 0.10 & 364$\pm$ 19$^h$\\
\enddata
\vskip 0.1in
$^a$ All parallaxes are from \citet{Kirkpatrick_2021}.\\
$^b$ $T_{\rm eff}$ is estimated from the photometry only, by calculating the average value determined from the color relationships of Table 1, for the adopted gravity and metallicity. The quoted uncertainty neglects any systematic errors in the models. \\
$^c$ For the WISE 0336 system, F480M is from \citet{Calissendorff_2023}, and F162M and F360M are estimated here, see Section 4. Quoted uncertainty is the sum of the squares of measurement error and color transformation uncertainty.\\
$^d$ The model sequences imply surface gravities for the primary and secondary components of the WISE 0336 system of log $g \approx 4.5$ and log $g \sim 4.0$ respectively, and solar metallicity. For the A component $T_{\rm eff}$ was estimated using Table 1 with both colors and $M_{F480M}$. For the B component the estimate uses $M_{F480M}$ only due to the divergence between the models and the observed F162M $-$ F480M color at the lowest temperatures (Figure 2).
The gravities and temperatures are consistent with a coeval system with an age of a few Gyr and a mass ratio around 2 \citep{Marley_2021}.\\
$^e$ For WISE 0359, magnitudes and uncertainties are from \citet{Beiler_2023}.\\
$^f$ The photometric temperature estimate adopts log $g = 4.5$ and solar metallicity. The result is consistent with the spectral fit presented in Section 3; the latter is the preferred value.\\
$^g$ For WISE 1828, magnitudes and uncertainties are from \citet{Furio_2023}.\\
$^h$ The photometric temperature estimate assumes that WISE 1828 is an equal-mass binary system, with log $g = 4.5$ and [m/H] $= -1.0$. The result is consistent with the spectral-fit temperature presented by \citet{Leggett_2021}.\\
\end{deluxetable*}

\subsection{WISE J033605.05-014350.4}

\citet{Calissendorff_2023} present {\em JWST} imaging observations of the Y0 dwarf WISE J033605.05-014350.4 (hereafter WISE 0336) in the F150W and F480M filters. The target was resolved into a binary system with a $0\farcs 09$ separation, where the secondary is  2 -- 3 magnitudes fainter than the primary.  This is a very low-mass system --- assuming an age of  1 -- 5 Gyr, evolutionary models indicate that the primary has a mass of 7.5 -- 20 Jupiter masses, while the secondary  has a mass of 4 -- 12.5 Jupiter masses \citep{Calissendorff_2023}.

We converted the \citet{Calissendorff_2023} WISE 0336A,B F150W magnitudes to F162M adopting the F150W $-$ F162M trend for Y dwarfs shown in Appendix C. For F360M we adopted the [3.6] magnitude of the unresolved system as that of the primary, given the large $\delta$mag between primary and secondary, and estimated F360M from that value.

WISE 0336A appears to be typical of the general population, based on its location in the color-color diagrams in Figure 2, and 
so the system is expected to have approximately solar metallicity and an age of a few Gyr. \citet{Calissendorff_2023} estimate $T_{\rm eff} = 415 \pm 20$~K for WISE 0336A,  by comparison to Sonora and ATMO2020++ model sequences. The relationships in Table 1 imply $T_{\rm eff} = 440 \pm 24$~K for the primary, assuming solar metallicity and log $g =4.5$, consistent with their value.


\citet{Calissendorff_2023} estimate $T_{\rm eff} = 325_{-10}^{+15}$~K for WISE 0336B.  Figure 2 suggests that the secondary is cooler, and the relationship in Table 1 gives $T_{\rm eff} = 295 \pm 10$~K, based on the F480M magnitude  only, and adopting log $g = 4.0$ and [m/H] $= 0$.  We find that WISE 0336B lies in the factor of two luminosity gap between WISE 0855 and the other Y dwarfs, which have $T_{\rm eff} \gtrsim 325$~K \citep[e.g.][]{Leggett_2021}. A resolved image in the F360M filter would be valuable for further analysis of this system.

Figure 2 suggests that WISE 0336A has log $g \approx 4.5$ and [m/H] $\approx 0.0$, and that WISE 0336B has a lower gravity. A difference in gravity for the two components of $\approx 0.5$~dex is consistent with a coeval system aged a few Gyr, with component temperatures of 440~K and 300~K, and a mass ratio of $\approx 2$ \citep{Marley_2021}. For example, a 3~Gyr old system composed of a 15 and a 7 Jupiter mass object would have ($T_{\rm eff}$, log $g$) values of (438, 4.6) and (295, 4.2), respectively.

The coolest known Y dwarf, the 260~K WISE J085510.83-071442.5 \citep{Luhman_2014, Leggett_2021}, hereafter WISE 0855, is highlighted in Figure 2.  
\citet{Leggett_2021} fit the WISE 0855 observations with parameters which differ from the standard grid values shown in the Figure, 
and the open stars illustrates the colors calculated by that model.
As also shown in Figure 9 of \citet{Leggett_2021}, 
Figure 2 indicates that the flux at F162M is underestimated by the model, but there is good agreement at 480M. It is intriguing that the \citet{Lacy_2023}
sequences with clouds show a significant bluewards trend in the F162M $-$ F480M colors. Although based only on a sample of two, the colors of WISE 0336B and WISE 0855 appear to be consistent with a disequilibrium log $g = 4.0$ atmosphere with water clouds, where the clouds are thinner than described by the E10 model of \citet{Lacy_2023}. It is also possible that further changes in the pressure-temperature profile, either as well as or instead of water cloud formation, could reproduce the colors of the coldest brown dwarfs. Additional {\em JWST} data will help clarify this issue.

\subsection{WISE J035934.06-540154.6}

\citet{Beiler_2023} calculate synthetic photometry for WISE 0359 using their observed spectrum (Section 3). Photometry for 27 {\em JWST} filters is provided in their Table 4. For WISE 0359 we converted the \citet{Beiler_2023} values in Jy to magnitudes using the magnitude to Jy values given for another Y dwarf by \citet{Furio_2023}.

WISE 0359 is typical of the general population, based on its location in the color-color diagrams in Figure 2 and our spectral analysis in the previous Section. We estimated 
$T_{\rm eff}$ from the F162M, F380M, and F480M colors using the log $g = 4.5$ solar metallicity relationships in Table 1. The value of $T_{\rm eff} = 463 \pm 28$~K
agrees well with the $T_{\rm eff} = 467^{+16}_{-18}$~K determined by Beiler et al. and the value we determined by analysis of the SED of  $T_{\rm eff} = 450$~K.

\subsection{WISEP J182831.08+265037.8}

\citet{Furio_2023} present {\em JWST} imaging observations of the Y dwarf 
WISEP J182831.08+265037.8  (hereafter WISE 1828), observed with NIRCam and NIRISS filters: F090W,	F115W,	F162M,	F335M,	F360M,	F470M,	and F480M. WISE 1828 is known to be unusual, being very bright and red in mid-infrared colors. The red colors suggest metal paucity \citep[Figure 2 and][]{Leggett_2021} and the brightness suggests that the system is multiple. However  De Furio et al. exclude a similarly bright companion at distances greater than 0.5 AU.
WISE 1828 continues to be very red and bright in  Figure 2. \citet{Leggett_2021} determine $T_{\rm eff} = 375$~K, log $g = 4.0$, and [m/H] $= -0.5$ from a spectral fit, assuming the system is an equal-mass binary. We estimated 
$T_{\rm eff}$ from the F162M, F380M, and F480M colors using the log $g = 4.5$ [m/H] $= -1$ relationships in Table 1, adopting the binary solution so that the value inferred from $M_{F480M}$ was consistent with that inferred from the other two colors. The value of $T_{\rm eff} = 364\pm 19$~K agrees well with the \citet{Leggett_2021} value.

\bigskip
\section{Conclusions} 

Since the discovery of what is now recognized as the first T dwarf, Gliese 229B, in 1995 \citep{Nakajima_1995}, it has been recognized that vertical transport of gas in turbulent brown dwarf atmospheres results in disequilibrium chemistry  \citep[e.g.][]{Fegley_1996, Griffith_1999, Lodders_1999}.

The Y dwarfs are now demonstrating that, as well as having atmospheres which are out of chemical equilibrium, the pressure-temperature ($P-T$) structure does not follow the standard radiative-convective adiabatic form \citep{Leggett_2021}.  This is not surprising, as these turbulent and fast rotating atmospheres are subject to dynamical, thermal, and chemical changes which disrupt the convective transport of heat from the lower to upper atmosphere \citep{Showman_2013, Tremblin_2015, Tan_2017, Showman_2019, Tremblin_2019, Tan_2021}.

All standard-adiabat atmospheric models for brown dwarfs cooler than 600~K generate SEDs which are too faint at wavelengths of 2~$\mu$m to 4~$\mu$m, and around 12~$\mu$m \citep[e.g.][]{Leggett_2021, Lacy_2023}. \citet{Leggett_2021} showed that a modified $P-T$ profile with a cooler lower atmosphere (where the near-infrared flux originates) and warmer upper atmosphere (where the mid-infrared flux originates)  improved the fit to observations substantially.  The discrepancy at $\lambda \approx 3.6~\mu$m, for example, was reduced by a factor of $\approx 5$ in the adiabat-adjusted models. Simulations of rapidly rotating cloud-free atmospheres (as generally applicable to Y dwarfs) calculate such a change in the $P-T$ profile  --- a cooler lower atmosphere and warmer upper atmosphere \citep[their Figure 4]{Tan_2021}. 

The first Y dwarf {\em JWST} observations to be published support the need for a non-adiabatic $P-T$ profile. 
We fit the first spectrum, of the Y0 dwarf WISE 0359 \citep{Beiler_2023}, with synthetic spectra generated by the ATMO2020++  adiabat-adjusted disequilibrium chemistry models. We find that these produce a superior fit, compared to standard-adiabat models.  The shape of the absorption features and the flux peaks are better reproduced. Furthermore the model generates a good fit without the need to use an atypical metallicity or age (surface gravity) for a field dwarf.  We strongly recommend that the ATMO2020++ models \citep{Phillips_2020, Leggett_2021, Meisner_2023} are used to analyse observations of brown dwarfs cooler than 600~K. Synthetic photometry and spectroscopy is available at 
\dataset[10.5281/zenodo.7931460] {https://zenodo.org/record/7931460}, and
\dataset[the ERC-ATMO Opendata page]
{https://noctis.erc-atmo.eu/fsdownload/Q7MUSoCLR/meisner2023}.

The {\em JWST} observation shows that the strong absorption by PH$_3$ expected at $\lambda \approx 4.3~\mu$m is not present in the Y dwarf spectrum. Spectra calculated by ATMO2020++ atmospheres with no phosphorus show a much improved agreement with the 
observations at  $4.1 \leq \lambda~\mu$m $\leq 4.3$. Pathways for phosphorus chemistry in turbulent  brown dwarf atmospheres with disequilibrium chemistry need to be reexamined.

We also explore the colors of four Y dwarfs observed with {\em JWST}, and recently published. We find that the two Y0 dwarfs, WISE 0359 and WISE 0336A, have colors typical of the field, and color-inferred temperatures which are consistent with earlier results. The recently discovered WISE 0336B \citep{Calissendorff_2023} is the first object to lie in the gap between the extreme WISE 0855 and all other Y dwarfs; we estimate that $T_{\rm eff} \approx 295$~K for this object. The fourth Y dwarf, WISE 1828 is known to be peculiar, and its {\em JWST} colors continue to indicate that it is an extremely metal poor binary system, although no companion has been found at separations greater than 5~AU \citep{Furio_2023}.

The locations of WISE 0336B and WISE 0855 in color-magnitude diagrams suggest that the ATMO2020++ models 
underestimate the flux at F162M and at F360M, for brown dwarfs with $T_{\rm eff} \lesssim 325$~K. It would not be surprising that the structure of the atmosphere changes for the coldest Y dwarfs, where  water clouds may form within the photosphere \citep[and Figure 2]{Lacy_2023}.  There may also be dynamical changes ---
\citet{Leggett_2021} find that the 260~K WISE 0855 is undergoing vigorous mixing with $K_{zz} = 8.7$, and that the deviation from a standard adiabat occurs deeper in the atmosphere at $P = 50$~bar, at temperatures where the nitrogen chemistry is changing and the chlorides and sulfides are condensing. 

We look forward to additional {\em JWST} data for the coldest Y dwarfs, which will enable exploration of this next frontier in the story of cold planetary-mass brown dwarfs.

\clearpage

\appendix

\section{ATMO2020++ Models: Additional Information}

The ATMO2020++ models include rainout of condensates which depletes refractory species, but they do not include clouds. The chemistry includes 277 species, and out-of-equilibrium chemistry was performed using the model of \citet{Tsai_2017}.

In the \citet{Leggett_2021} study of the SEDs of seven brown dwarfs, values  of $1.2 \leq \gamma \leq 1.33$ were found 
for the upper-atmosphere adiabat. Values of  $7 \leq P(\gamma ,\max ) $~bar $\leq 15$ were found for 
the pressure of the layer below which the adiabat increases to the standard value (around 1.4 for a diatomic gas). Exceptions were the late-T dwarf where  $P(\gamma ,\max )$ was undefined, and the extremely cold 260~K dwarf, where $P(\gamma ,\max) = 50$ bar.  Interestingly, for the Y dwarfs the values of $T$ at  $P(\gamma ,\max )$ are all  $\sim 800$~K, a temperature where the nitrogen chemistry is changing and chlorides and sulfides are condensing, processes which may disrupt convection. 

The larger grid of \citet{Meisner_2023} generalized the atmospheric parameters 
in the following ways:
\begin{itemize}
    \item The levels with a modified adiabat are between 0.15 and 15 bars at log $g = 4.5$ and are scaled by $\times {10}^{\mathrm{log}(g) - 4.5}$ at other surface gravities. For these layers a value of $\gamma = 1.25$ is adopted. Higher and lower layers use the standard adiabat for that layer, typically a value around 1.4 (the ratio of specific heats for the gas at that layer).
    \item Out-of-equilibrium chemistry is used with $K_{zz} = 10^5$ cm$^2$ s$^{-1}$ at log $g = 5.0$ which is scaled by $\times {10}^{2(5-\mathrm{log}(g))}$ at other surface gravities. 
    \item The mixing length is assumed to be 2 scale heights at 1.5 bars, and higher pressures, at log $g = 4.5$, and is scaled down by the ratio between the local pressure and the pressure at 1.5 bars for lower pressures. The 1.5 bars limit is scaled by $\times {10}^{\mathrm{log}(g)-4.5}$ at other surface gravities. 
\end{itemize}

The ATMO2020++ synthetic Y dwarf SEDs are similar to those calculated by the earlier \citet{Leggett_2021} models, except at $YJH$ ($1.0 \leq \lambda~\mu$m $\leq 1.5$) where the ATMO2020++ flux is smaller 
by $\sim 0.5$~magnitudes. We trace this to the  parametrization
of the mixing length, which leads to stronger convective fluxes that reduce the temperatures in the deep atmosphere, where the near-infrared flux originates. The agreement with observations of T and Y dwarfs remains good, as can be seen in Figures 2 and 3 of \citet{Meisner_2023}, and Section 3 of this work.

\begin{figure}[hb!]
\plotone{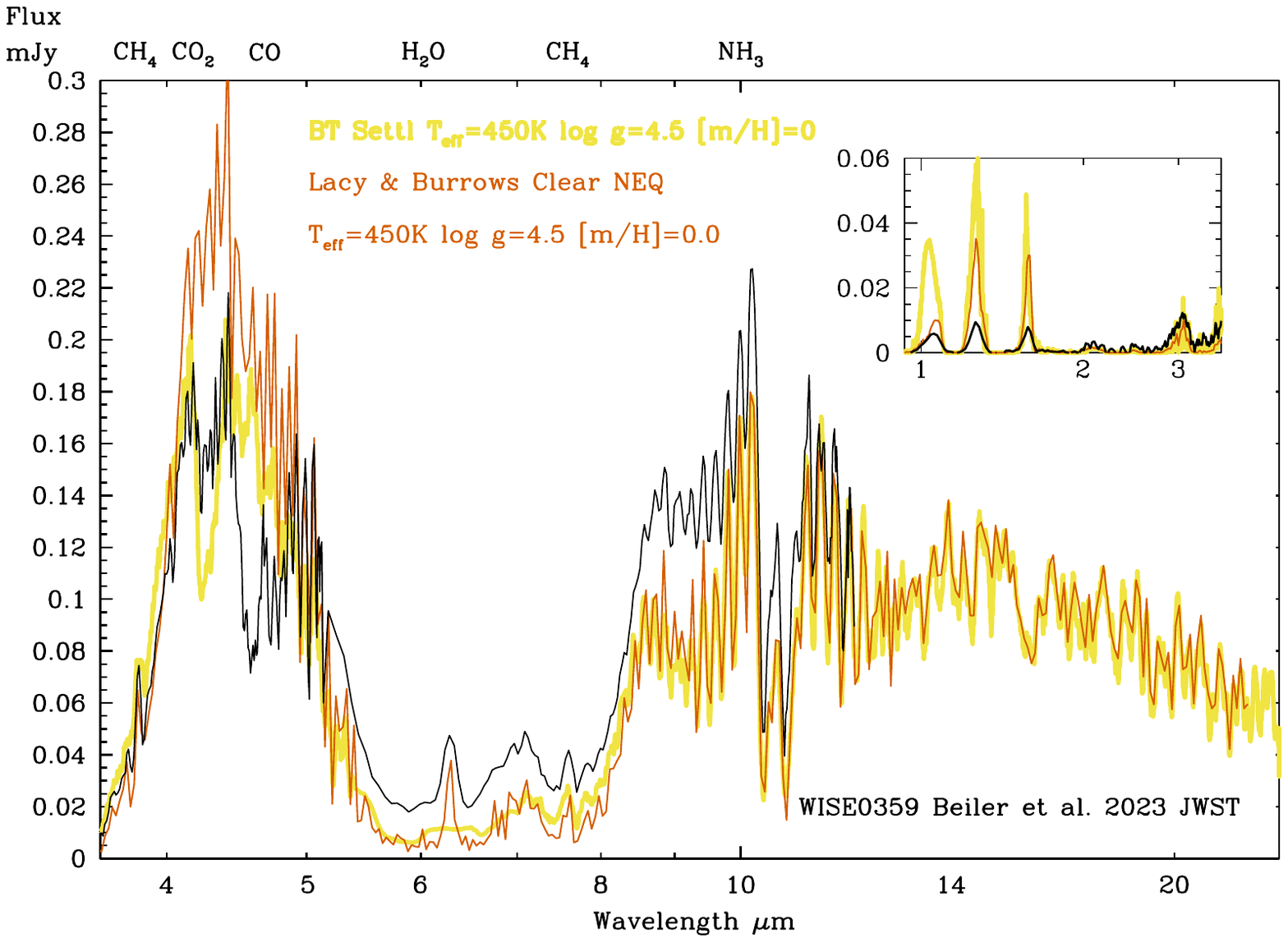}
\vskip -0.2in
\caption{The black line is the observed spectrum of  WISE 035934.06-540154.6 from \citet{Beiler_2023}.  Yellow and orange lines are synthetic spectra generated by model atmospheres with 
$T_{\rm eff} = 450$~K, log $g = 4.5$, and solar metallicity from the BT-Settl suite \citet{Allard_2014, Allard_2016} and by \citet{Lacy_2023}, respectively.
Principal opacity species are indicated along the top axis.}
\end{figure}

\section{Comparison to BT-Settl and Lacy \& Burrows Synthetic Spectra}

Figure 4 compares the observed spectrum for WISE 0359 to synthetic spectra generated by BT-Settl \citep{Allard_2014, Allard_2016} and \citet{Lacy_2023} model atmospheres. These models produce an inferior fit compared to the Sonora suite and ATMO2020++ (Figure 1), at these temperatures.

\section{Color Transformations}

\begin{figure}[ht!]
\plotone{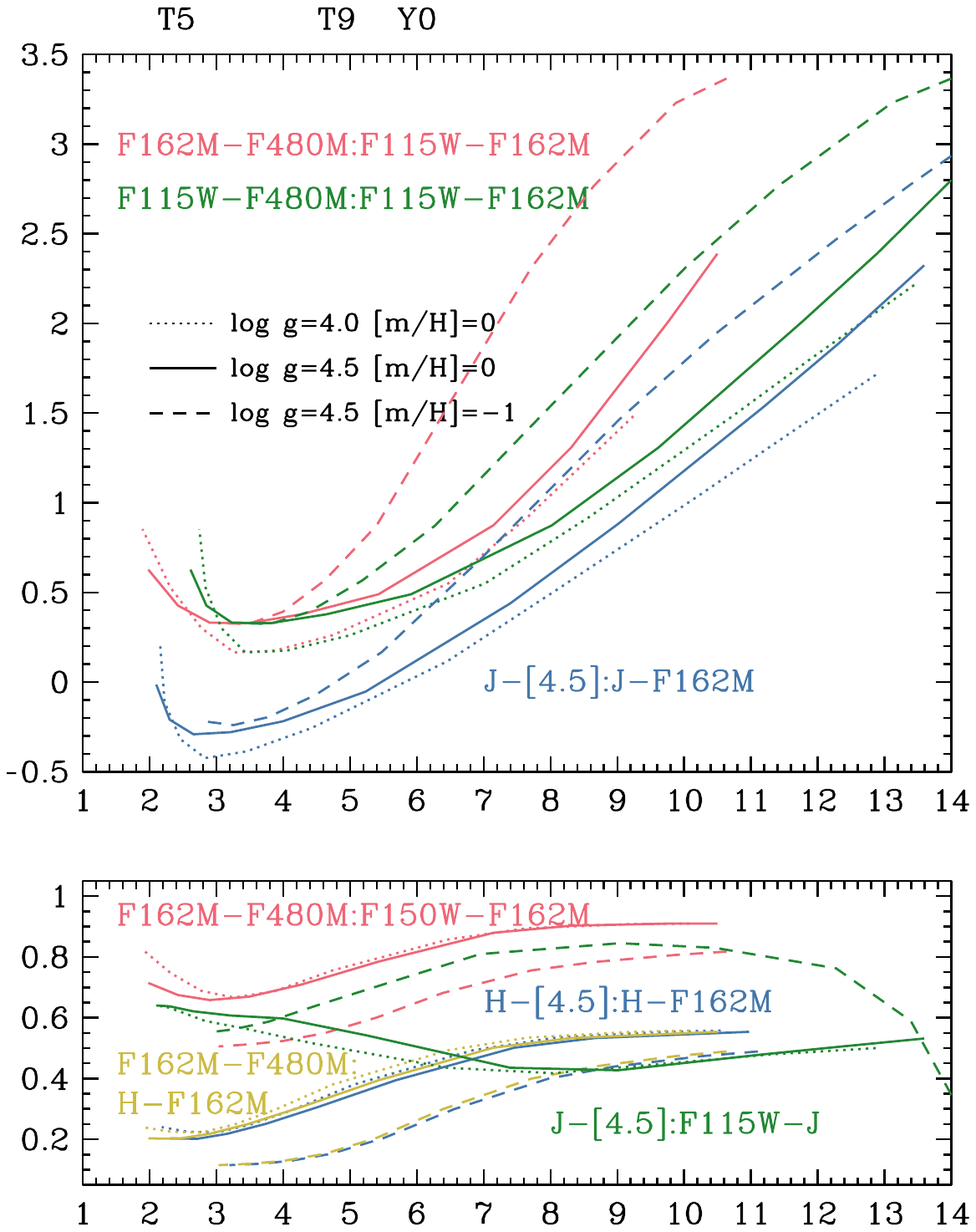}
\vskip -0.3in
\caption{Color transformations calculated from the ATMO2020++ grid, with surface gravity and metallicity as in the legend.  Colors corresponding to $x:y$ are identified in the colored legends. Approximate spectral types are indicated along the top axis.
}
\end{figure}

Figures 5 and 6 show relationships between {\em JWST} colors and those of ground based, {\em Spitzer} and {\em WISE}. These have been calculated using the ATMO2020++ models.  Table 3 gives polynomial relationships between the colors.

\begin{figure}[hb!]
\plotone{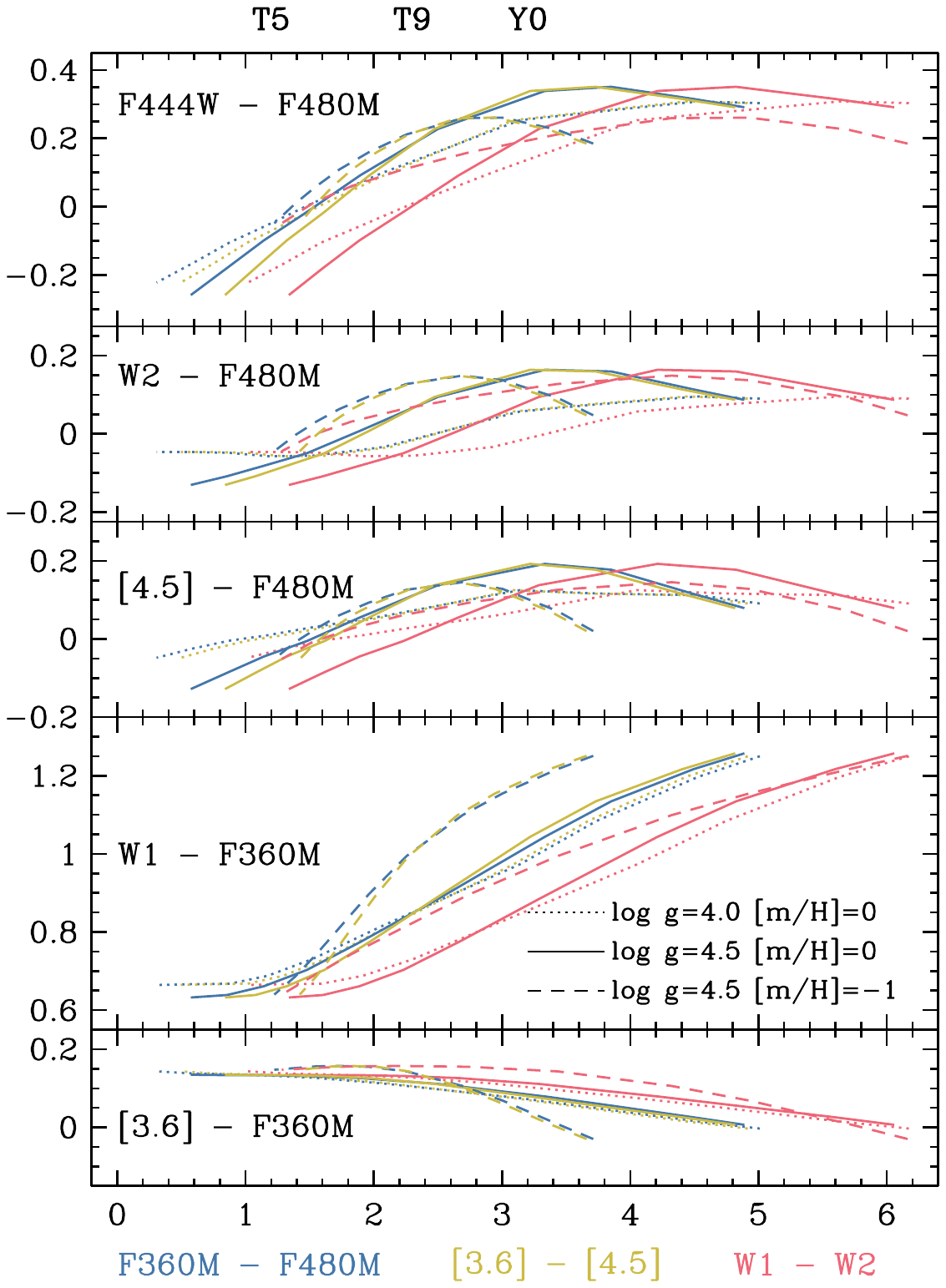}
\vskip -0.2in
\caption{Additional color transformations for longer wavelengths, calculated from the ATMO2020++ grid, with surface gravity and metallicity as in the legend.  The color corresponding to $x$ is indicated by the color of the sequence and identified along the $x$-axis. The color corresponding to $y$ is identified in each panel.  Approximate spectral types are indicated along the top axis.
}
\end{figure}

\begin{longrotatetable}
\centerwidetable
\begin{deluxetable*}{cclllllllllllllll}
\setlength{\tabcolsep}{3pt}
\tabletypesize{\scriptsize}
\tablecaption{Polynomial Fits to Color Transformations}
\tablehead{
\colhead{x} & \colhead{y} &  \multicolumn{5}{c}{log $g=4.0$ [m/H]$=$0.0}  &  \multicolumn{5}{c}{log $g=4.5$ [m/H]$=$0.0}   &  \multicolumn{5}{c}{log $g=4.5$ [m/H]$=$-1.0} \\
   &      & \colhead{$a_0$} & \colhead{$a_1$} & \colhead{$a_2$}
& \colhead{$a_3$} & \colhead{$a_4$}  & \colhead{$a_0$} & \colhead{$a_1$} & \colhead{$a_2$}
& \colhead{$a_3$} & \colhead{$a_4$}  & \colhead{$a_0$} & \colhead{$a_1$} & \colhead{$a_2$}
& \colhead{$a_3$} & \colhead{$a_3$} 
} 
\startdata
F115W$-$F480M & F115W$-$F162M &
4.673 & -2.5526 & 4.8497e-1 & -3.6058e-2  & 9.7454e-4  &
2.989 & -1.4773 & 2.6780e-1 & -1.8117e-2  & 4.6809e-4  &
1.982 & -9.7842e-1 & 1.7616e-1 & -8.3881e-3  & 9.0558e-5 \\
$J -$ [4.5] & F115W $- J$ &
0.681  & 1.3380e-2  & -2.2707e-2  & 2.9787e-3 &  -1.0723e-4 & 
0.478  & 1.5643e-1 &  -4.8441e-2  & 4.6474e-3 &  -1.4042e-4 & 
0.203  & 2.1398e-1 &  -4.3699e-2 &  4.9428e-3 & -2.0384e-4 \\
$J -$ [4.5] & $J -$  F162M &
1.388  &  -1.2195  &  2.7135e-1  &  -2.1488e-2  &  6.1572e-4   & 
1.042  &  -8.9331e-1  &  1.9192e-1  &  -1.3854e-2  &  3.7402e-4   & 
1.592  &  -1.2442  &  2.6855e-1  &  -1.8811e-2  &  4.6147e-4  \\
F162m $-$ F480M & F115W $-$ F162M &
3.776  & -2.4432 &  5.6448e-1  & -5.2695e-2 &  1.8680e-3 & 
2.498  & -1.4454  &  3.1351e-1  & -2.6196e-2 &  8.8862e-4 & 
2.740 &  -1.6434  & 3.2747e-1  & -1.6011e-2  &  2.3314e-5 \\
F162m $-$ F480M & F150W - F162M &
1.599 &  -6.9319e-1  & 1.7604e-1   & -1.7107e-2 &  5.7365e-4 & 
1.248 &  -4.7626e-1 &  1.2752e-1  &  -1.2514e-2  & 4.1829e-4 & 
0.787  & -2.5357e-1 &  6.9023e-2   & -6.2225e-3 & 1.8822e-4\\
F162m $-$ F480M &  $H -$ F162M & 
0.548  & -3.2733e-1  & 1.0484e-1  & -1.1100e-2  & 3.8914e-4 & 
0.424 &  -2.3810e-1 &  7.9194e-2 &  -8.2098e-3  & 2.7955e-4 & 
0.493 &  -3.0877e-1  & 7.7394e-2  & -6.4643e-3  & 1.8014e-4\\
$H -$ [4.5] &  $H -$ F162M & 
0.641 &  -3.7627e-1  & 1.0968e-1 &  -1.0940e-2 &  3.6457e-4 & 
0.510 &  -2.9149e-1 &  8.7509e-2 &  -8.6120e-3  & 2.8124-4 & 
0.570  & -3.5161e-1 &  8.5301e-2  & -7.1288e-3 &  2.0174e-4\\
F360M $-$ F480M  & [3.6] $-$ F360M & 
0.154 &  -2.8668e-2 & 1.3874e-2   &  -5.8890e-3 & 6.0052e-4 &
0.141 &  -9.1162e-3  & 6.5372e-3 &  -4.1263e-3 & 4.1047e-4 &
0.188 &  -1.6226e-1 & 1.7889e-1 & -6.8038e-2 & 7.3584e-3 \\
F360M $-$ F480M &  W1 $-$ F360M & 
0.650  & -1.3559e-2  & 5.2231e-2  & -4.0487e-3  & -2.0583e-4 & 
0.640  & -8.6658e-2 &  1.1520e-1 &  -1.8654e-2 &  8.2271e-4 & 
0.829  & -8.5596e-1 &  8.2946e-1  & -2.3589e-1  & 2.2272e-2\\
F360M $-$ F480M &  [4.5] $-$ F480M & 
-0.067 &  5.7680e-2 &  1.4777e-2 &  -6.1914e-3 &  4.5013e-4 & 
-0.174 &  2.1521e-2 &  1.1162e-1 &  -3.4872e-2 &  2.7251e-3 & 
-0.313 &  9.2330e-2 &  1.8543e-1 &  -7.9976e-2  & 8.0523e-3\\
F360M $-$ F480M &  W2 $-$ F480M & 
-0.038 &  -3.5574e-2  & 1.2764e-2 &  5.8833e-3 &  -1.2033e-3 & 
-0.128 &  -9.1847e-2 &  1.5610e-1  & -4.1067e-2 &  3.0365e-3 & 
-0.303 &  1.0963e-1 &  1.5884e-1  & -7.0166e-2 &  7.0907e-3\\
F360M $-$ F480M  & F444W $-$ F480M & 
-0.269 &  1.5542e-1  & 4.1496e-2 &  -1.6725e-2 &  1.3613e-3 & 
-0.372  & 1.4207e-1  & 1.2363e-1 &  -4.2174e-2  & 3.3990e-3 &  
-0.465	 & 2.5418e-1 &  1.5006e-1 &  -7.4212e-2 &  7.5719e-3\\
{[}3.6] $-$ [4.5]  & [3.6] $-$ F360M & 
0.1601  & -3.6315e-2  & 1.8091e-2  & -7.1210e-3 & 7.2268e-4 &
0.140  & -5.2979e-4  & 6.0450e-4  & -2.9935e-3 & 3.4957e-4 &
0.158 &  -1.4516e-1  & 1.9005e-1  & -7.6667e-2 & 8.6544e-3\\
{[}3.6] $-$ [4.5]  &  W1 $-$ F360M & 
0.659  & -5.3340e-2 &  7.5974e-2  & -8.3955e-3  & 3.4069e-5 & 
0.707  & -2.6021e-1  & 2.2554e-1  & -4.3652e-2  & 2.6950e-3 & 
1.187  & -1.6904 &  1.3999 &  -3.8879e-1 &  3.6507e-2\\
{[}3.6] $-$ [4.5]  &  [4.5] $-$ F480M & 
-0.080 &  5.3854e-2 &  2.3359e-2  & -8.9039e-3 &  6.8505e-4 & 
-0.170  & -7.7409e-2 &  1.9398e-1 &  -5.7038e-2  & 4.6299e-3 & 
-0.546 &  2.4352e-1  & 1.9550e-1  & -1.0130e-1 &  1.1279e-2\\
{[}3.6] $-$ [4.5]  &  W2 $-$ F480M  & 
-0.029 &  -7.2903e-2 &  4.9725e-2  & -6.4368e-3  & 8.7387e-5 & 
-0.072  & -2.5604e-1  & 2.6740e-1  & -6.8545e-2 &  5.2865e-3 & 
-0.538 &  2.7810e-1 &  1.5150e-1 &  -8.5494e-2 &  9.6464e-3\\
{[}3.6] $-$ [4.5] &  F444W $-$ F480M & 
-0.305 &  1.4997e-1  & 5.9916e-2  & -2.2247e-2 &  1.8287e-3 & 
-0.412  & 4.5887e-2 &  2.2360e-1 &  -7.0240e-2  & 5.8296e-3 & 
-0.831	 & 5.4561e-1  & 1.0461e-1  & -8.5415e-2 &  1.0066e-2\\
W1 $-$ W2 &  [3.6] $-$ F360M & 
0.178 &  -4.7478e-2 &  2.0570e-2 &  -5.3062e-3 & 3.9592e-4 &
0.142 &  -4.5846e-3 &  3.5817e-3  & -2.1292e-3 & 1.7279e-4 &
0.129  & 7.3250e-3  & 1.1340e-2  & -4.6950e-3 & 3.2163e-4\\
W1 $-$ W2  & W1 $-$ F360M & 
0.6839  &  -8.5829e-2  & 5.7446e-2  & -4.5123e-3 &  1.8382e-5 & 
0.7882  & -3.1417e-1  & 1.7673e-1  & -2.6207e-2 &  1.7208e-3 & 
0.3578  & 2.3486e-1 &  -1.0831e-2 &  -1.6181e-3 &  1.6374e-4\\
W1 $-$ W2  &  [4.5] $-$ F480M & 
-0.101  & 3.6900e-2  & 2.0078e-2 &  -5.2343e-3 &  3.0048e-4 & 
-0.152  & -1.3719e-1  & 1.5776e-1 &  -3.4133e-2 &  2.1229e-3 & 
-0.459 & 4.5278e-1  & -1.3427e-1  & 1.9382e-2  & -1.2047e-3\\
W1 $-$ W2  &  W2 $-$ F480M & 
0.001 &  -6.0762e-2 &  1.1390e-2  & 4.1738e-3 & -6.5666e-4 & 
0.016 &  -3.4203e-1 &  2.2428e-1 &  -4.2757e-2   & 2.5365e-3 & 
-0.438  & 4.3977e-1  & -1.3197e-1  & 1.9292e-2  &  -1.1909e-3\\
W1 $$-$$ W2 &  F444W $-$ F480M & 
-0.360 &  9.9529e-2 &  5.4674e-2  & -1.4146e-2 &  8.9198e-4 & 
-0.453  & -1.4270e-2 &  1.7170e-1 &  -4.0272e-2 &  2.5825e-3 & 
-0.634 & 	6.5764e-1 &  -2.0223e-1 &  3.0507e-2 &  -1.8585-3\\
\enddata
\tablecomments{
$y$ is estimated using\\
$y = a_0 + a_1\times{x} + a_2\times{x}^2  + a_2\times{x}^3 + a_4\times{x}^4$\\
Residuals in  $y$ are  typically $\pm 0.1$ mag.
\\
}
\end{deluxetable*}
\end{longrotatetable}


\bibliography{JWST_dY}{}
\bibliographystyle{aasjournal}

%



\end{document}